\documentclass[12pt,osajnl2,preprint,showpacs]{revtex4}
\DeclareRobustCommand{\baselinestretch{2.2}}

\usepackage{graphicx}

\begin{document}
\title{Nonlinear photonic crystals near the supercollimation point}
\author{Zhiyong Xu$^{1,2}$, Bj\"{o}rn Maes$^{3}$, Xunya Jiang$^{4}$, John D. Joannopoulos$^{1}$,
Lluis Torner$^{2}$ and Marin Solja\v{c}i\'{c}$^{1}$}
\affiliation{$^{1}$Dept.\ of Physics, Research Laboratory of
Electronics, Massachusetts Institute of Technology, Cambridge
02139, Massachusetts, USA} \affiliation{$^{2}$ICFO-Institut de
Ciencies Fotoniques, and Universitat Politecnica de Catalunya,
Mediterranean Technology Park, 08860 Castelldefels (Barcelona),
Spain} \affiliation{$^{3}$Photonics Research Group, Ghent
University, St.-Pietersnieuwstraat 41, 9000 Ghent, Belgium}
\affiliation{$^{4}$Institute of Microsystem and Information
Technology, CAS, Shanghai 200050, People's Republic of China}
\address{$^*$Corresponding author: xzy124@rsphysse.anu.edu.au}

\begin{abstract}

We uncover a strong coupling between nonlinearity and diffraction
in a photonic crystal at the supercollimation point. We show this
is modeled by a \emph{nonlinear diffraction term} in a nonlinear
schr\"{o}dinger type equation, in which the properties of solitons
are investigated. Linear stability analysis shows solitons are
stable in an existence domain that obeys the Vakhitov-Kolokolov
criterium. In addition, we investigate the influence of the
nonlinear diffraction on soliton collision scenarios.
\end{abstract}

\maketitle
Photonic crystals (PhC) are under active investigation due to the
rich physics and all-optical signal control\cite{Joannopoulos,Yablonovitch,John}. One
of the striking features in PhCs is the supercollimation (SC) effect,
that was experimentally\cite{Kosaka,Pustai} and
theoretically\cite{Notomi,Witzens,Shin} examined. This effect
originates from the possibility of obtaining flat regions in the
spatial dispersion relation or equifrequency contours of PhCs. At
these particular points the phase propagation components along the
direction of a beam are equal. In this way all
components of the beam travel with the same phase velocity and
it becomes nondiffractive. Recently, centimeter-scale
SC was achieved in a large-area two-dimensional
PhC\cite{Rakich}. However, the study of nonlinear effects around
these SC points is largely unaddressed.
Here we propose a more fundamental approach, as opposed to a more
numerical one\cite{Jiang}.

In this letter we derive a phenomenological model that describes
light beam propagation in nonlinear PhCs around the
SC point. In most optical systems, nonlinearity is a
small perturbation, and hence added to the linear equation of
motion. In contrast, in the current system, one
uncovers a strong coupling between nonlinearity and diffraction.
More specifically, a nonlinear diffraction term is introduced into
the nonlinear Schr\"odinger (NLS) equation. This term can control
the magnitude, and sometimes even the sign of the diffraction. As
a result, this is a unique system, and new physical phenomena
emerge. For example, this modified equation gives rise to
solitons up to an upper threshold for the propagation constant. Linear
stability analysis demonstrates the solitons are stable in the
existence domain, and obey the Vakhitov-Kolokolov criterium. In
addition, we examine soliton interaction scenarios.

\begin{figure}[t]
\begin{center}
\includegraphics[width=6cm,height=6cm, bb=184 509 377 700]{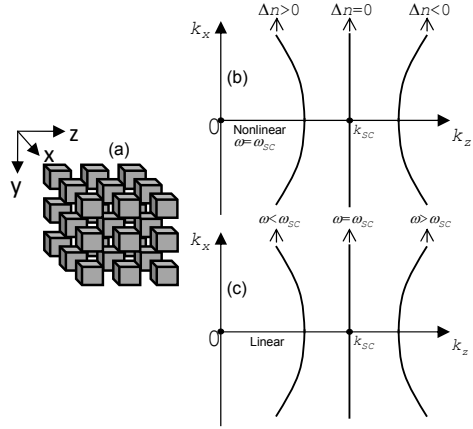}
\end{center}
\caption{(a) A scheme of 3D PhC. (b) Depiction of the linear dispersion relation in the
proximity of a SC point. (c) Nonlinearity gives rise
to an index change $\delta n$, and is modeled by a shift of the
dispersion relation, here shown for the particular input frequency
$\omega=\omega_{SC}$.} \label{figure0}
\end{figure}

In order to deduce the model equations, we consider a 3D PhC with
the same periodicity in the transversal $x$ and $y$ directions (a scheme shown in Fig.~\ref{figure0}), and
write the electric field of a beam in the PhC as ${\bf E}(x,y,z,t)
={\bf F}(x,y,z)A(x,y,z)\exp(ik_z^c z-i\omega t)$, with $k_z^c$ the
propagation constant of the central Bloch mode, $\omega$ the
frequency of the beam, $A(x,y,z)$ the slowly varying amplitude,
and ${\bf F}(x,y,z)$ the Bloch mode profile corresponding to
$\omega$ and $k_z^c$. The propagation direction is along $z$.

We study beams in the neighborhood of an SC point,
so $\omega\approx\omega_{SC}$ with $\omega_{SC}$ the SC frequency. The special
dispersion relation around this point is shown in Fig.~\ref{figure0}. The sign and strength of diffraction
depends strongly on the position of $\omega$ versus $\omega_{SC}$.
Exactly at the frequency $\omega_{SC}$, all components of the beam travel
with the same propagation constant along $z$, noted by $k_z^{SC}$, and there is no diffraction.
For frequencies different than $\omega_{SC}$ we approximate the equifrequency contours by a
parabola. In this case diffraction can be positive or negative, depending on the position of
$\omega$ versus $\omega_{SC}$. Furthermore, the strength of diffraction increases as the difference between
$\omega$ and $\omega_{SC}$ increases.

To include nonlinearity we apply first order
perturbation theory\cite{Marin}. Thus, the nonlinear interaction
causes a small shift of the local dispersion relation [Fig.~\ref{figure0}], which is equivalent to
shifting $\omega_{SC}$. To first order this shift
is given by
\begin{eqnarray}
&&\frac{\Delta \omega}{\omega_{SC}}=-\frac{1}{4}\frac{\int \mathrm{d}{\bf r}
\mathrm{ n}_2({\bf r})\mathrm{n}({\bf r})[({\bf F\cdot F})({\bf
F}^{*}{\bf \cdot F}^{*})+2\left| {\bf F}\right| ^4]}{\int
\mathrm{d}{\bf r}\mathrm{n}^2({\bf r})\left|
{\bf F}\right| ^2} \nonumber \\
&&\times \left| A(x,y,z)\right| ^2 \equiv \kappa \left| A(x,y,z)\right|
^2
\end{eqnarray} with $\kappa$ a nonlinear coefficient calculated from the linear Bloch mode
profile.

The nonlinearity shifts $\omega_{SC}$ to
$\omega_{SC}+\Delta\omega$, so by Taylor expanding the dispersion relation
around the SC point, the modified dispersion relation becomes approximately
\begin{eqnarray}\label{eq:disp}
k_z-k_z^{SC}+\alpha_1(\omega -\omega_{SC}-\kappa \left|
A\right|^2\omega_{SC})\nonumber \\
=\beta_1(\omega -\omega_{SC} -\kappa \left|
A\right|^2\omega_{SC})(k_x^2+k_y^2)
\end{eqnarray} with $k_z$ the longitudinal,
and $k_x$, $k_y$ the transverse propagation vector components. The term with
$\alpha_1$ corresponds to the frequency dependence of the central
$k_z$-component (thus at $k_x,k_y=0$). Similarly, the term with
$\beta_1$ describes the frequency change of the curvature (or the
diffraction). Note that we neglect higher-order diffraction terms,
such as fourth-order diffraction, as we are considering
propagation of a broad beam with respect to the PhC period.
We transform Eq.~\ref{eq:disp} into the space domain, and arrive
at the following equation in dimensionless form: $i\frac{\partial
q}{\partial \xi}+\frac 12\alpha \nabla^{2}q -\frac 12\beta \left|
q\right|^2\nabla^{2}q+\gamma \left|q\right|^2 q=0$,
where
$\nabla^{2}=\frac{\partial^{2}}{\partial\eta^{2}}+\frac{\partial^{2}}{\partial\zeta^{2}}$,
with the transverse coordinates $\eta$ and $\zeta$ scaled to the
spatial characteristic width $W_0$ and $\xi$ is the longitudinal
coordinate scaled to the free-space diffraction length $L_d=2 \pi
W_0^2/\lambda$, for Gaussian-like beams\cite{Lax}. $\alpha=-2
\beta_1(\omega-\omega_{SC})L_d/W_0^2$ indicates the linear
diffraction strength, and its sign characterizes the
type of linear diffraction. The novel nonlinear diffraction term
is preceded by $\beta=2\beta_1\kappa\omega_{SC} c^2 L_d/W_0^2$,
with $c$ the speed of light. The usual nonlinear term has
$\gamma=-\alpha_1\kappa\omega_{SC} c^2 L_d$. In the following, we
employ $\delta n(I)>0$ (thus $n_2>0$) and $\beta_1>0$.

In this paper we will mainly show results concerning 2D PhCs, so the
resulting equation takes the form
\begin{equation}
\label{eq:model} i\frac{\partial q}{\partial \xi}+\frac{1}{2}\alpha
\frac{\partial^{2}q}{\partial \eta^{2}} -\frac{1}{2}\beta \left|
q\right|^2\frac{\partial^{2}q}{\partial \eta^{2}}+\gamma
\left|q\right|^2 q=0.
\end{equation}
Note that all the coefficients for this model can be deduced
from rigorous numerical simulations \cite{Rakich}. Recently, an equation similar to Eq.~\ref{eq:model} was reported
as the continuous approximation of the Salerno
model\cite{Salerno,Gardenes1,Marklund}. Here in contrast, we
derive the model from a physical system, describing light beam
propagation in nonlinear PhCs with SC. Eq.~\ref{eq:model} conserves the power
$U=-\frac{1}{\beta} \int \ln \left| \alpha -\beta \left| q\right|
^2\right|\mathrm{d}\eta$.
The stationary solutions of Eq.~\ref{eq:model} have the form
$q(\eta,\xi)=w(\eta) \exp(ib \xi)$, where a real
function $w(\eta)$ and a real propagation constant $b$ are found by iterative relaxation method. To analyze
stability we examine perturbed solutions $q=(w+u+iv) \exp(ib
\xi)$, where the real $u(\eta,\xi)$ and imaginary $v(\eta,\xi)$
perturbations can grow with complex rate $\delta$ upon
propagation. The eigenvalue problem linearized from Eq.~\ref{eq:model} around $w(\eta)$ is solved numerically. 

\begin{figure}[t]
  \begin{center}
  \includegraphics[width=7cm, bb=189 360 425 576]{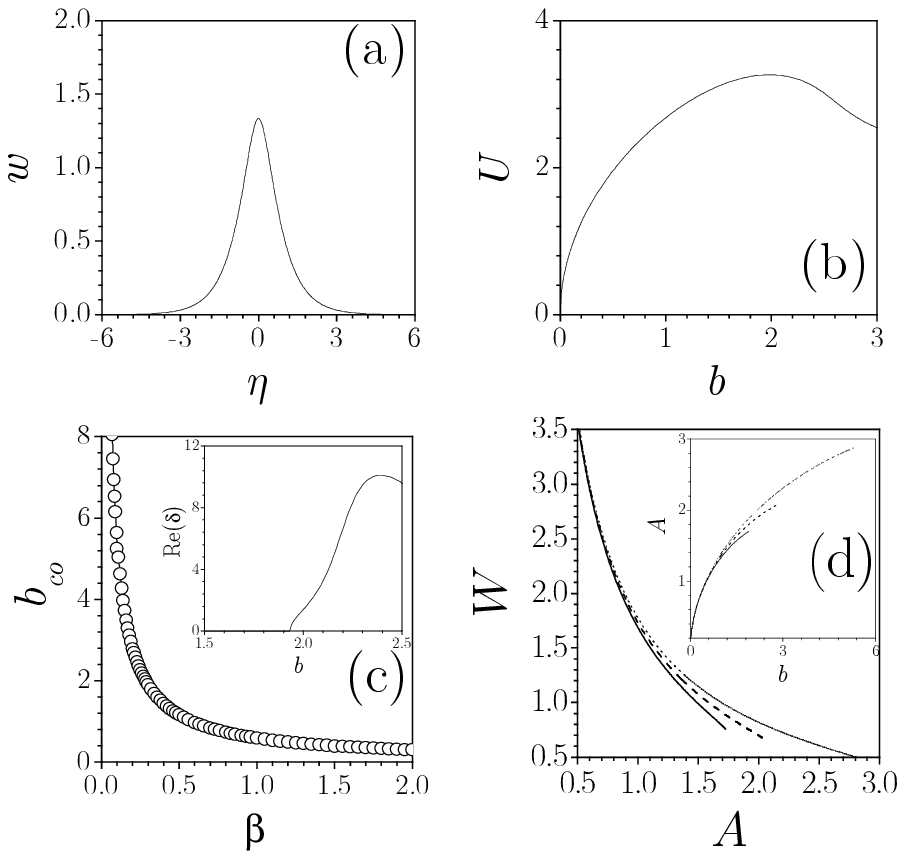}
  \end{center}
  \caption{(a) Profile of a soliton for $b=1$ with $\beta=0.3$.
  (b) The dispersion diagram for $\beta =0.3$.
  (c) Domain of existence for solitons in the ($b,\beta$) plane, where
  the inset shows the real part of the perturbation growth rate versus
  propagation constant for $\beta =0.3$. (d) The soliton width (FWHM) versus maximum amplitude for varying nonlinear diffraction,
  namely $\beta=0.1$ (dotted line), $0.2$ (dashed
  line) and $0.3$ (solid line). The inset shows the dependence of maximum amplitude on the propagation
constant. $\alpha=1$ and $\gamma=1$ for all cases. }
  \label{figure1}
\end{figure}

One concludes that Eq.~\ref{eq:model} allows soliton solutions
[see e.g.~Fig.~\ref{figure1}(a)], but only in a finite band of
propagation constants. More specifically, there exists an upper
threshold for the propagation constant ($b_{co}$), which depends
on the strength of nonlinear diffraction $\beta$
[Fig.~\ref{figure1}(c)]. Above this value peakon solutions\cite{Gardenes1} appear,
which are unphysical given the assumptions of our physical model.
To model what happens beyond that point in a real PhC, one would
have to include more terms to Eq.~\ref{eq:model}. To emphasize,
in our system it is \emph{the nonlinear diffraction term} that leads to
the reduction of the semi-infinite band of propagation constants
(as in a pure cubic NLS system) to a finite one, which is typical for
soliton families in models with competing nonlinearities, such as
the cubic-quintic NLS equation\cite{Pushkarov}. It is noted that
the power is a non-monotonic function of the propagation
constant [Fig.\ref{figure1}(b)]. Fig.\ref{figure1}(d) shows the
dependence of the width of solitons on the amplitude (inset plot
shows the dependence of maximum amplitude on the propagation
constant). As one can see from this plot, the nonlinear
diffraction has a significant effect at larger amplitudes. In
addition, it should be pointed out that a lower power would be
needed to generate a soliton in our system because the diffraction
is so weak to start with. Thus our system may provide an experimentally favorable
way to manipulate nonlinear waves.

Linear stability analysis shows that solitons are stable in the whole
domain of their existence. An instability growth rate calculation is shown in the inset of
Fig.~\ref{figure1}(c). It is noted that the Vakhitov-Kolokolov
criterium applies to our system for fundamental solitons, namely
solitons are stable when $\frac{\mathrm{d}U}{\mathrm{d}b}>0$.
To confirm the outcome of the linear stability analysis we perform
direct numerical simulations of Eq.~\ref{eq:model}. We employ a
split-step Fourier method, in combination with fourth-order
Runge-Kutta to deal with the nonlinear diffraction term.
The input condition is $q(\eta,\xi=0)=w(\eta)[1+\rho(\eta)]$, with
$w(\eta)$ the profile of the stationary soliton and $\rho(\eta)$ a
random noise function with variance up to $10\%$. The simulations
confirm the linear stability analysis.

\begin{figure}[t]
\begin{center}
\includegraphics[width=6cm, bb=221 518 399 710]{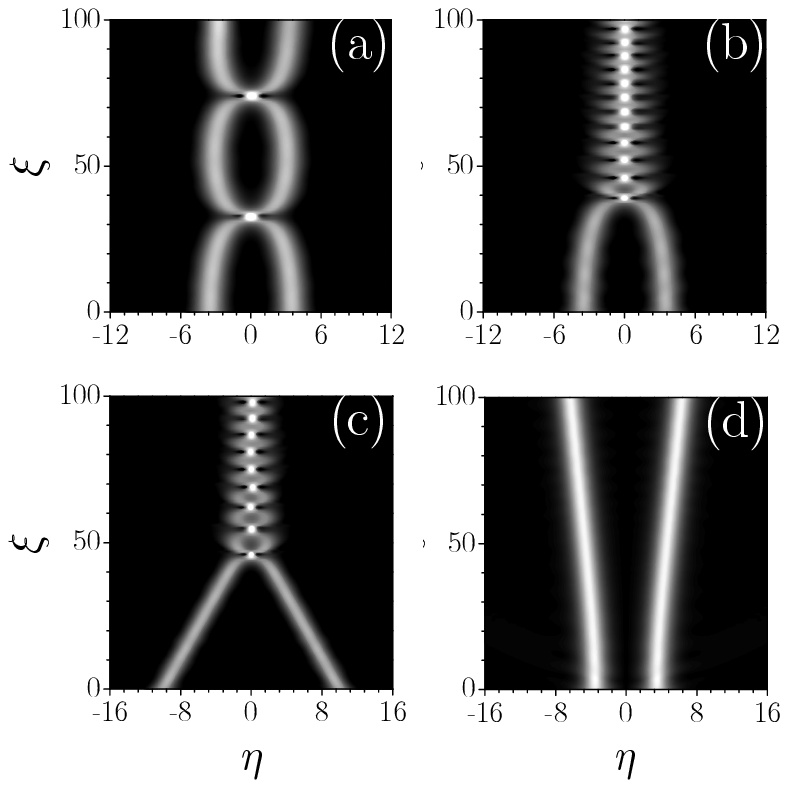}
\end{center}
\caption{Collision scenarios between solitons with
$\alpha=\gamma=1$: Two in-phase parallel solitons for (a)
$\beta=0.05$ and (b) $\beta=0.15$. (c) Two solitons moving in
opposite directions with an angle ($\theta=0.2$), and $\beta=0.15$.
(d) Two out-of-phase parallel solitons
for $\beta =0.15$.} \label{figure3}
\end{figure}

As another example where nonlinear diffraction affects fundamental
phenomena, we report its influence on soliton collisions for
different input conditions. We use as input a soliton of
the pure NLS ($\beta=0$), and examine how changing
$\beta$ affects collision scenarios. In the case of two
in-phase parallel solitons as input, the colliding solitons behave
periodically for a small nonlinear diffraction term
[Fig.~\ref{figure3}(a)], as in the case of the pure NLS. For
larger $\beta$ the colliding solitons merge into a single
localized state, with breather-like features
[Fig.~\ref{figure3}(b)]. For solitons moving in opposite
directions, the colliding solitons feature similar behavior. One
example, presented in Fig.~\ref{figure3}(c), shows two solitons
merging into a single localized state. Such inelastic effects often
appear when dealing with a perturbed NLS, again, here it is caused purely
by nonlinear diffraction. For two
out-of-phase solitons, the results show that the repulsive force
between neighboring solitons is reduced with an increase of
nonlinear diffraction [Fig.~\ref{figure3}(d)].


Finally, we investigated soliton properties in the (2+1)D model
(two transversal dimensions and one propagation direction) but no
stable solitons were found. The results show however that nonlinear
diffraction modifies the critical power for collapse: positive (negative) $\beta$
reduces (increases) the critical power. This means the nonlinear diffraction
can slow down the collapse. 

In most PhC structures the SC effect is relatively broadband.
This means that the curvature changes slowly near the SC
point. Therefore, to have a significant nonlinear diffraction effect one
needs to design a PhC with a large curvature change, meaning a large $\beta_1$.
At the SC frequency the nonlinear diffraction term would then be the
main contribution that can interact with the normal nonlinear term.


This work was partially supported by the Institute for Soldier
Nanotechnologies under Contract No. W911NF-07-D-0004. BM
acknowledges support from the Funds for Scientific
Research-Flanders (FWO-Vlaanderen). The authors acknowledge the
support of the Interconnect Focus Center, one of five research
centers funded under the Focus Center Research Program, a DARPA
and Semiconductor Research Corporation program.

\pagebreak
\renewcommand\refname{\textbf{References with titles}}

\pagebreak
\renewcommand\refname{\textbf{References without titles}}

\pagebreak
\textbf{Figure captions}

\textbf{Figure 1}. (a) A scheme of 3D PhC. (b) Depiction of the linear dispersion relation in the
proximity of a SC point. (c) Nonlinearity gives rise
to an index change $\delta n$, and is modeled by a shift of the
dispersion relation, here shown for the particular input frequency
$\omega=\omega_{SC}$..

\textbf{Figure 2}. (a) Profile of a soliton for $b=1$ with $\beta=0.3$. (b) The dispersion diagram for $\beta =0.3$.
(c) Domain of existence for solitons in the ($b,\beta$) plane, where the inset shows the real part of the perturbation growth rate versus propagation constant for $\beta =0.3$. (d) The soliton width (FWHM) versus maximum amplitude for varying nonlinear diffraction, namely $\beta=0.1$ (dotted line), $0.2$ (dashed line) and $0.3$ (solid line). The inset shows the dependence of maximum amplitude on the propagation constant. $\alpha=1$ and $\gamma=1$ for all cases.

\textbf{Figure 3}. Collision scenarios between solitons with
$\alpha=\gamma=1$: Two in-phase parallel solitons for (a)
$\beta=0.05$ and (b) $\beta=0.15$. (c) Two solitons moving in
opposite directions with an angle ($\theta=0.2$), and $\beta=0.15$.
(d) Two out-of-phase parallel solitons
for $\beta =0.15$.

\pagebreak

\pagebreak

\pagebreak

\end{document}